\documentclass{article}

\usepackage{arxiv}
\usepackage{amsmath}
\usepackage{amssymb}
\usepackage{xcolor}
\usepackage{graphicx}

\usepackage[utf8]{inputenc} 
\usepackage[T1]{fontenc}    
\usepackage{hyperref}       
\usepackage{url}            
\usepackage{booktabs}       
\usepackage{amsfonts}       
\usepackage{nicefrac}       
\usepackage{microtype}      
\usepackage{lipsum}

\usepackage{bm}

\newcommand{\norm}[1]{\left\lvert \left\lvert #1 \right\lvert \right\lvert}

\title{The Gene Mover's Distance:\\Single-cell similarity via Optimal Transport}

\author{
Codegoni Andrea, Gualandi Stefano, Vercesi Eleonora\\
Department of Mathematics ``F. Casorati'', University of Pavia, Italy\\
{\tt \{andrea.codegoni01, eleonora.vercesi01\}@universitadipavia.it, stefano.gualandi@unipv.it}\\

\And
Bellazzi Riccardo, Nicora Giovanna\\
Department of Bioengineering, University of Pavia, Italy\\
{\tt riccardo.bellazzi@unipv.it, giovanna.nicora01@universitadipavia.it}\\
}

\begin{document}
\maketitle

\begin{abstract}
This paper introduces the Gene Mover's Distance, a measure of similarity between a pair of cells based on their gene expression profiles obtained via single-cell RNA sequencing.
The underlying idea of the proposed distance is to interpret the gene expression array of a single cell as a discrete probability measure.
The distance between two cells is hence computed by solving an Optimal Transport problem between the two corresponding discrete measures.
In the Optimal Transport model, we use two types of cost function for measuring the distance between a pair of genes.
The first cost function exploits a gene embedding, called {\tt gene2vec}, which is used to map each gene to a high dimensional vector: the cost of moving a unit of mass of gene expression from a gene to another is set to the Euclidean distance between the corresponding embedded vectors.
The second cost function is based on a Pearson distance among pairs of genes.
In both cost functions, the more two genes are correlated, the lower is their distance.
We exploit the Gene Mover's Distance to solve two classification problems: the classification of cells according to their condition and according to their type.
To assess the impact of our new metric, we compare the performances of a $k$-Nearest Neighbor classifier using different distances.
The computational results show that the Gene Mover's Distance is competitive with the state-of-the-art distances used in the literature.

\end{abstract}

\keywords{Optimal Transport \and Network Simplex \and Single-Cell RNA Sequencing \and $k$-Nearest neighbour}

\section{Introduction}\label{sec1}
In Computational Biology, the recent advances in single-cell RNA sequencing have produced a large amount of high dimensional raw data that has to be analyzed in order to extract meaningful information \cite{Kim2019}.
The main output of a single cell RNA Sequencing experiment is a gene expression profile.
The expression profile is obtained after the sequence of the transcriptome: the genome sequence of a cell tells us what the cell could do, while the expression profile tells us what it is doing at a given moment, that is when the sequencing experiment has done.
Thus, a gene expression profile is the pattern of the expressed genes under specific circumstances or in a type of cell.
These profiles can be used, for instance, to distinguish between cells that are performing their normal functions and cells that are not, or show how cells react to a particular treatment.
Best practices to analyze raw count matrices of gene expression are based on statistical inference tools.
After preprocessing the raw data, a typical analysis consists of clustering the cells based on the similarity of the gene expression profiles of every single cell.
Gene expression profiles are given as columns of the so-called count matrices, where there is a row for each gene and a column for each cell.
While the number of cells (columns) is only up to a few thousand, the number of genes is larger than $20\,000$.
However, most of the genes are not expressed in every cell, and, hence, the columns of the count matrices are represented by very sparse vectors, that is, most of the elements of gene expression arrays are zero entries.
Note that there is a strong analogy between gene expressions and the text documents, where a {\it bag of words} representations yield very sparse vectors \cite{Pekalska2001}.
Due to the very high dimension of gene expression profiles, along with their sparsity, the raw count matrices are currently processed with dimension reduction algorithm, such as, for instance, Principal Component Analysis (PCA) or the $t$‐distributed Stochastic Neighbour Embedding ($t$-SNE) \cite{Luecken2019}.
However, as remarked in \cite{Kim2019}, the core concept underlying any clustering algorithm is the metric used as similarity measure.


In this work, we propose a novel application of Computational Optimal Transport to measure the similarity between a pair of cells using the Gene Mover's Distance.
%
Intuitively, the idea is to measure the similarity between a pair of cells by solving an Optimal Transport problem that minimizes the cost of transforming the gene expression profile of the first cell into the gene expression profile of the second cell.
The cost of this transformation depends on the cost of moving a single unit of mass (of gene expression) from one gene to another.
This latter cost, in the framework of Optimal Transport, is called the {\it ground distance}.
If two gene expressions are equal, we do not move any unit of mass, and clearly, the overall cost is zero.
Otherwise, the cost of transforming a gene expression into the other is given by the sum over all the displacements of units of mass (of gene expression).
Notice that to compute the distance between a pair of genes we have to solve a Linear Programming problem, which can be efficiently solved using the Network Simplex algorithm (e.g., see \cite{Gualandi2020}).

In the Gene Mover's Distance, we have tested two types of ground distances.
The first ground distance exploits the {\tt gene2vec} embedding recently introduced in \cite{Du2019} which maps each single gene into a high dimensional vector.
In {\tt gene2vec}, two genes that are highly correlated have two embeddings, that are two vectors of $\mathbb{R}^{200}$, which have a small Euclidean distance.
The second ground distance that we use is based on a Pearson distance matrix which is computed time by time, depending on the cells involved in the experiment.
In the Pearson distance matrix, pairs of genes which are strongly correlated have a small distance.
%

%
%
%
%
%
%
%

Optimal Transport has emerged in the last decade as a powerful mathematical tool to analyse and compare high dimensional data via different variants of the Wasserstein distance \cite{villani2008optimal,santambrogio2015optimal,Peyre2019}.
The computer vision community has applied Optimal Transport metrics since the late '90, via the so-called {\it Earth Mover's Distance} to compare or classify images, to perform point set registration, and to implement adaptive color transfer \cite{Rubner2000,Pele2009,Bonneel2019,Rabin2014}.
In statistics and probability, the Wasserstein distance is known as the Mallow's distance \cite{Bickel2001}, and it is used to assess the goodness of fit between distributions \cite{Munk1998,Sommerfeld2018} as well as an alternative to the usual $g$-divergences as a cost function in minimum distance point estimation problems \cite{Bassetti2006,Bassetti2006b}.
In computational biology, we mention the application to flow cytometry diagrams \cite{Orlova2016}, in gene expression cartography \cite{Nitzan2019}, and the identification of developmental trajectories in reprogamming \cite{schiebinger2019optimal}.
A very successful application of Optimal Transport is the {\it Word Mover's Distance} used to compare text documents \cite{WMD2015}.
The Word Mover's Distance has directly motivated our work, since there is a strong analogy with the sparse representation of text documents into a high dimensional space, via the exploitation of the embedding of words into $\mathbb{R}^{300}$ given by {\tt word2vec} \cite{Mikolov2013} or {\tt GloVe} \cite{Glove2014}.

The main contributions of this paper are the following.
\begin{enumerate}
    \item We introduce the Gene Mover's Distance (GMD), a new measure of (dis)similarity between cells, which leverages recent results on the embedding of gene expressions into a high dimensional space, using the {\it gene2vec} embedding of \cite{Du2019}.
    \item We apply the GMD distance to (i) a binary classification problem to distinguish between {\it normal} and {\it malignant} cells in patients affected by Acute Myeloid Leukemia (AML), and to (ii) to a multi-class classification problem for pancreatic and brain cells, to distinguish among different types of cells of the same tissue.
  Both classification problems are based on a $k$-Nearest Neighbour classifiers, which is used to compare four different distance functions: GMD+{\tt gene2vec}, GMD+{\tt Pearson}, the Euclidean distance, and the Pearson correlation score among gene expressions.

    \item We propose a simplified SC-RNA dataset to promote the research in single-cell classification based on gene expression profiles\footnote{Our dataset is published online at \url{https://zenodo.org/record/4604569}}. This dataset uses a format similar to other Machine Learning dataset, and it should encourage other researchers to enter the SC-RNA application domain \cite{zenodo2021}.
%
\end{enumerate}

\paragraph{Outline.} The outline of this paper is as follows.
Section~\ref{cell:background} introduces the main features of single-cell RNA sequencing and gene expression data analysis.
Section~\ref{sec:GMD} presents the Linear Programming formulation of the Gene Mover's Distance which is solved by the Network Simplex algorithm.
Section~\ref{sec:data} describes in detail the biological datasets and the preprocessing we apply to the data.
Section~\ref{sec:comp} reports our computational results based on biomedical data for Acute Myeloid Leukemia, pancreatic, and brain cells.
We conclude the paper in Section~\ref{sec:concl} with a perspective on future works.

\section{Single-Cell Sequencing and Data Analysis}\label{cell:background}
Recent improvements in high-throughput sequencing technologies allow the measurement of a patient's transcriptomic profiles at single-cell resolution, thus  providing opportunities to improve diagnosis and to individuate treatments based on cellular heterogeneity \cite{hedlund_single-cell_2018}.
The transcriptome, i.e. the set of RNA molecules produced by a cell, is an indicator of the gene expression (GE), which in turn identifies the state of the cell at a given time and under certain circumstances  \cite{hwang_single-cell_2018}.
Several approaches have been developed to analyse single-cell data and to classify each cell according to its type, based on their gene expression profile \cite{abdelaal_comparison_2019}.
Furthermore, as discussed in \cite{Luecken2019}, thanks to Single Cell RNA Sequencing (SC-RNA) experiments, scientists have been able to study cell heterogeneity in different species, discovering previously unknown cell types.
SC-RNA Sequencing has been largely used to split up cells into different types and sub-types (e.g., see the survey \cite{darmanis2015survey}).

Finding the link between different cells type trough different organisms is important from a medical perspective.
As pointed out in \cite{baron2016single}, a clinical trial may be done on model organisms, such as mice or monkeys, with far less ethical problems, and then transposed to humans.
Moreover, some tissues, such as the brain, have a wide and non fully understood complexity.

Cancer could greatly benefit from the application of SC-RNA Sequencing since this disease arises and progresses from a population of (often heterogeneous) malignant cells \cite{ellsworth_single-cell_2017}.
Several efforts have been made in the last years to characterize different cancer types at a single-cell resolution \cite{10.1093/nar/gky939}.
For instance, these studies have revealed the clonal architecture and heterogeneity, especially in hematological cancer \cite{potter_single_2019, fan_linking_2018, hou_single-cell_2012, hughes_clonal_2014}.
The ability to detect Circulating Tumor Cells (CTCs) on blood samples could boost cancer diagnosis, as well as supporting clinical decision making, for instance by tailoring treatment according to the cancer genome \cite{rossi_single-cell_2019}.
Acute Myeloid Leukemia (AML) is a blood cancer characterized by the development of abnormal cells in myeloid lineage that interfere with normal blood cells.
This causes tiredness and easy bleeding and increases the risk of infections.
This condition is worsened by heterogeneous responses to treatments: this is why ~75\% of patients relapse within 5 years of diagnosis \cite{van2019single}.
There are several risk factors for leukemia (for instance smoke and Down Syndrome, see \cite{lagunas2017acute} for an overview), but it is well known that certain changes in DNA can transform some normal cells into malignant ones.
Furthermore, it is important to distinguish among acquired DNA changes versus inherited ones.
Accordingly to these differences, there are different subgroups of AML and each of them can have different changes in genes and chromosomes.
For a full comprehension of the concept briefly explained above, the reader can refer to the website of the American Cancer Society \cite{ACS}.

The actual application of SC-RNA sequencing in cancer clinical practice demands reliable computational tools that efficiently analyse a large amount of data, and convert them into actionable knowledge.
In particular, SC gene expression analysis involves several steps since such data are inherently noisy with confounding factors, for instance,  biological and technical variables \cite{hwang_single-cell_2018}.

An important step forward in providing a functional description of gene expression profiles is the idea of finding distributed representation of genes using a gene embedding called {\tt gene2vec}, in the same spirit of the word embedding, such as the {\tt word2vec} \cite{Mikolov2013} or {\tt GloVe} \cite{Glove2014}.
In \cite{Du2019}, the authors propose to map every single gene appearing in one of the 984 selected datasets from the GEO database, into a continuous Euclidean space, in such a way that co-expressed genes that appear in the same context are mapped into ``close'' Euclidean vectors.
While the authors of {\it gene2vec} introduced their embedding to predict gene-gene interactions, in the next section, we show how we can exploit theirs embedding to define a new measure of similarity between single-cell that exploits the gene co-expression learned by the embedding.

\section{The Gene Mover's Distance}
\label{sec:GMD}

In this section, we describe the Gene Mover's Distance in terms of a Linear Programming problem that can be efficiently solved by the Network Simplex algorithm.
The input data consists of two vectors of gene expression associated with two cells.
The number of genes for each cell is up to $30\,000$, but only a tiny fraction of these genes have strictly positive values, that is, the gene expressions are very sparse vectors.


%

Let $\mathcal{G}$ denote the set of every possible gene, and let $\mu$ and $\nu$ denote two vectors of gene expression profiles associated with two different cells $c_\mu$ and $c_\nu$.
In our work, we consider only the genes whose expression is strictly positive.
Hence, the two vectors do not necessarily have the same size, and they might have different genes expressed.
Let us denote by $I \subseteq \mathcal{G}$ the set of genes positively expressed in cell $c_\mu$, and by $J \subseteq \mathcal{G}$ the set of genes expressed in cell $c_\nu$.
Since we want to interpret the two gene expression vectors as discrete probability measures, we normalize the two vectors $\mu$ and $\nu$ to sum up to 1, as follows:
\begin{equation}\label{sumone}
\begin{array}{rcl}
    \bar \mu_h &:=& \frac{\mu_h}{\sum_{i \in I} \mu_i}, \quad \forall h \in I, \\
    \bar \nu_h &:=& \frac{\nu_h}{\sum_{j \in J} \nu_j}, \quad \forall h \in J.
\end{array}
\end{equation}
\noindent Using $\mathcal{P}(\mathbb{R}^n)$ to indicate the set of probability measures over $\mathbb{R}^n$, we have that $\bar \mu \in \mathcal{P}(\mathbb{R}^{|I|})$ and $\bar \nu \in \mathcal{P}(\mathbb{R}^{|J|})$.
To represent the cost of moving a unit of gene expression from gene $i \in I$ to gene $j \in J$, we introduce the cost function
$C:\mathcal{G}\times \mathcal{G} \to \mathbb{R}^+$. In Optimal Transport terms, this is called the {\it ground distance}.
Since we are in a discrete setting, this cost function has a matrix form.

Given two cells $c_\mu$ and $c_\nu$ with normalized gene expressions $\bar \mu$ and $\bar \nu$, respectively, and given a ground distance $C$,
we define the Gene Mover's Distance between $c_\mu$ and $c_\nu$ as the optimal solution value of the following Linear Programming problem:
\begin{align}
\label{bgmd:eq1}
\mbox{GMD}(c_\mu, c_\nu) := \Bigl(\min_{\pi \in \mathcal{P}(\mathbb{R}^{I\times J})} \;
& \sum_{i\in I} \sum_{j\in J} C(i,j)^2 \,\pi_{ij}\Bigr)^{\frac{1}{2}} \\
    \mbox{s.t.} \quad
    & \sum_{j \in J} \pi_{ij} = \bar \mu_i, & \forall i \in I, \\
    & \sum_{i \in I} \pi_{ij} = \bar \nu_j, & \forall j \in J, \\
\label{bgmd:ver}    & \pi_{ij} \geq 0, & \forall i \in I, \forall j \in J.
\end{align}
This problem is an instance of the classical Kantorovich's Optimal Transport problem \cite{Kantorovich1960}, and it can be formulated as an uncapacitated network flow problem on a bipartite graph.
In practice, as we will discuss in Section~\ref{sec:comp}, it is efficiently solved by our implementation of a specialized Network Simplex algorithm, which exploits the geometry of the symmetric cost function.

In problem \eqref{bgmd:eq1}--\eqref{bgmd:ver}, the cost function $C$ has a fundamental role.
The theory of Optimal Transport guarantees that whenever $C:\mathcal{G}\times \mathcal{G} \to \mathbb{R}^+$ is a {\it metric}, that is, a distance function which satisfies the axioms of identity, symmetry, non-negativity, and the triangle inequality, then also GMD is a metric \cite{villani2008optimal}.
By taking the square of $C(i,j)$ and by computing the square root of the optimal value, we get a Wasserstein distance of order 2.
Hence, the GMD inherits all the mathematical properties of Wasserstein distances. We refer the reader to \cite{santambrogio2015optimal} an in-depth overview of Wasserstein distances.

In this paper, we consider two different metrics for the ground distance $C(i,j)$ appearing in \eqref{bgmd:eq1}.
The first metric is based on an embedding of the genes in $\mathbb{R}^n$, the second is based on a particular Pearson distance matrix.
In \cite{Du2019}, the authors have computed an embedding of genes into $\mathbb{R}^{200}$, called {\tt gene2vec}, by looking for the embedding minimizing the distances among highly functional correlated genes.
The vectors of $\mathbb{R}^{200}$ are obtained by training a deep neural network that takes as input the information about the co-expression of genes in functional patterns.
The underlying idea, borrowed by the {\tt word2vec} embedding \cite{Mikolov2013}, is that two genes that are frequently co-expressed end up having a small Euclidean distance between their two embedded vectors.
%
More formally, we can define the embedding as the function {\tt gene2vec}$: \mathcal{G} \rightarrow \mathbb{R}^{200}$, which maps any gene $i\in \mathcal{G}$ to a vector $\bm x_i \in \mathbb{R}^{200}$.
Using this embedding, we can define the ground distance:
\begin{equation}\label{cost:g2v}
    C^{\mbox{g2v}}(i, j) := \norm{\bm x_i - \bm x_j}_2,
\end{equation}
where $\bm x_i$ and $\bm x_j$ are the images through {\tt gene2vec}$: \mathcal{G} \to \mathbb{R}^{200}$ of $i, j \in \mathcal{G}$ respectively.
Clearly, $C^{\mbox{g2v}}$ is a metric.
%

The second ground distance we have considered is based on a Pearson distance matrix, which is precomputed as follows.
Let $\mathcal{C}$ be the set of all the cells contained in a given dataset.
We fix a subset of cells $\bar{\mathcal{C}} \subseteq \mathcal{C}$ and we compute the {\it Pearson correlation coefficient} $\rho_{i,j}$ between $i,j \in \mathcal{G}$ as follows.
\begin{equation}\label{eq:pearson}
    \rho_{i,j} \; := \;
    \frac{\displaystyle\sum_{\mu \in \bar{\mathcal{C}}} (\mu_i -  m_i)  (\mu_j -  m_j)}
    {\sqrt{\displaystyle\sum_{\mu \in \bar{\mathcal{C}}} (\mu_i -  m_i)^2 \, \sum_{\mu \in \bar{\mathcal{C}}} (\mu_j -  m_j)^2}},
\end{equation}
where $m_i$ and $m_j$ are the sample means in $\bar{\mathcal{C}}$ of genes $i$ and $j$, respectively.
Then, we set
\begin{equation}\label{cost:pearson}
 C^P(i,j) := \sqrt{1-\rho_{i,j}}.
\end{equation}
When two genes are positively correlated this distance function yields small values.
Notice the $C^P(i,j)$ is an axiomatic Pearson metric thanks to the square root, as proven in \cite{Deza2009}.
However, this cost function depends on the choice of the subset of cells $\bar{\mathcal{C}}$ used in \eqref{eq:pearson}.
In our test, we have used all cells for each given dataset.
%
%

Currently, the similarity metrics most used in the biomedical literature are the Euclidean distance and Pearson similarity score \cite{Kim2019,Luecken2019}.
Given the two gene expression profiles $\mu$ and $\nu$, they are defined as
\begin{enumerate}
\item Euclidean distance:
\begin{equation}\label{dist:eucl}
    d_E(c_\mu, c_\nu) := \norm{\mu - \nu}_2=
    \sqrt{\sum_n (\mu_n-\nu_n)^2   }.
\end{equation}
\item The Pearson similarity score:
\begin{equation}\label{dist:pears}
\hspace{-0.7cm}
    d_P(c_\mu, c_\nu) := 1 - \frac{\displaystyle\sum_{h \in H} (\mu_{h} -  m_\mu)  (\nu_{h} -  m_\nu)}{\sqrt{\displaystyle\sum_{h\in H} (\mu_{h} - m_\mu)^2 \, \sum_{h \in H} (\nu_{h} -  m_\nu)^2}},
\end{equation}
\noindent where $m_\mu$ and $m_\nu$ denote the sample mean of the two gene expression profiles and $H \subseteq \mathcal{G}$ is the subset of genes expressed in both cells. As noted in \cite{Deza2009}, the distance in \eqref{dist:pears} induces only a semi-metric, since the triangle inequality does not hold.
\end{enumerate}

We stress that \eqref{dist:pears} is different from \eqref{eq:pearson}: the first assumes that each $\mu_h$ is a sample of the gene expression ``representative'' of cell $\mu$ while $h$ varies.
The second considers each $\mu_h$ as a sample of the actual value of the gene expression of gene $g_h$ as $\mu$ varies.
If we consider a matrix that has all the $\mu \in \bar{\mathcal{C}}$ as columns, \eqref{dist:pears} computes the correlation between columns, while \eqref{eq:pearson} between rows.
%

\section{Datasets and Preprocessing}\label{sec:data}
We use three different open datasets published by the National Center for Biotechnology Information \cite{NCBI}: Acute Myeloid Leukemia (GSE116256, \cite{van2019single}), human pancreas (GSE84133, \cite{baron2016single}), and human brain (GSE67835, \cite{darmanis2015survey}).
Table \ref{tab:riassuntodataset} reports the main features of the three datasets.
The first column gives the unique access number of each dataset (accessible online at \cite{NCBI}).
The other columns, for each dataset, report the cell tissue, the number of patients used to collect the data, the overall number of cells, labels, classes, and genes.

Next, we first describe every single dataset, and then we discuss two important preprocessing steps.
The datasets used in this paper are published online on Zenodo \cite{zenodo2021}.
While the original data available on the Gene Expression Omnibus (GEO) database are spread on several different files, with a large amount of additional records, our format is very similar to other Machine Learning datasets, where all the relevant information are contained into a single easy to parse file.

\subsection{Datasets description}\label{sec:dataset}
We have considered the following three datasets.
\begin{table}[t!]
\centering
\caption{Main features of the  dataset.\label{tab:riassuntodataset}}
\begin{tabular}{llrrrrrr}
\hline
    Dataset & Tissue & Patients & Cells & Labels & Classes & Genes\\
\hline
    GSE116256 & Blood  & $15$  & $38\,410$ & $2$ & $12$ & $27\,899$\\
    GSE84133 & Pancreas & $4$ & $8\,569$ & $1$  & $14$ & $20\,124$\\
    GSE67835 & Brain & $12$ & $466$ & $1$ & $9$ & $22\,083$ \\
\hline
\end{tabular}
\end{table}

\begin{itemize}
    \item {\bf GSE116256}:
    Acute Myeloid Leukemia (AML) is a blood cancer characterized by the development of abnormal cells in myeloid lineage that interfere with normal blood cells \cite{van2019single}.
    This causes tiredness and easy bleeding and increases the risk of infections.
    This condition is worsened by heterogeneous responses to treatments: this is why $\sim75\%$ of patients relapse within 5 years of diagnosis.
    Due to this heterogeneity, different subgroups of AML exist, and each subgroup can have large differences in genes and chromosomes: SC-RNA Sequencing can help in detecting and studying such kinds of subpopulations, and, as a consequence, to enable advances in precision medicine \cite{seow2020single}.
    In \cite{van_galen_single-cell_2019}, authors use an elaborated supervised procedure to distinguish malignant and normal cells in 16 AML patients and 5 healthy donors.
    The resulting dataset consists of the gene expression profiles of $39\,000$ cells, which are labeled by their condition (normal/malignant) and type (Monocyte, Dendritic cells, \ldots).
    We refer to \cite{van_galen_single-cell_2019} for the biomedical details about the original raw data of this dataset.

    \item {\bf GSE84133}:
    The human pancreas is made up of 14 different types of cells: mast, endothelial, beta, epsilon, macrophage, acinar, gamma, quiescent stellate, T-cell, schwann, ductal, alpha, delta, activated stellate \cite{baron2016single}.
    The dysfunction of the pancreas are clinically important, as they are strictly related to several diseases, such as type 1 diabetes (T1D) and cancer.
    A lot of efforts have tried to replace lost Beta cells in T1D patients, but the main obstacle is the lack of understanding at the molecular level of the cell types.
    In \cite{baron2016single}, the authors aim to detect new biological properties of pancreatic cells starting from SC-RNA Sequencing experiments.
    They perform a classification by type using hierarchical clustering on cells from four human donors and two mice strains.
    In our test, we consider only the human samples.

    \item {\bf GSE67835}: This dataset concerns human brain cells. The brain is a tissue extremely complex, and it is composed of multiple cell classes \cite{darmanis2015survey}.
    Furthermore, it is hard to obtain brain samples to perform any type of analysis, since the human brain can be analyzed only {\it post mortem}.
    For these reasons, as in the pancreas, it is important to lead studies on non human brains.
    The authors of \cite{darmanis2015survey} have collected and analyzed trough SC-RNA Sequencing 8 adult and 4 fetal brain samples to assess both the diversity between cells of the adult brain and between pre- and post-natal ones.
    They are able to identify 10 clusters: 8 clusters are identified by, first, running a hierarchical clustering, and, then, by gene enrichment analysis into one cell type among Oligodendrocytes Precursor Cells (OPCs), oligodendrocytes, astrocytes, microglia, neurons, endothelial cells, replicating neuronal progenitors, and quiescent newly born neurons.
    The remaining two clusters are labeled \emph{hybrids}, as they contain genes characteristic of different cell types. 
\end{itemize}

In the NCBI database \cite{NCBI}, these three datasets are decomposed into several files, with different formats, and they contain information which is irrelevant for our research.
In order to encourage further research in single-cell analysis, we will release a simplified version of the datasets, where for each cell we report the fundamental labeled data.
Our dataset format closely resembles the structure of other standard benchmarks used by the Machine Learning community.
Our simplified datasets will be available online in case of acceptance of this paper.

\subsection{Gene Expression Normalization}\label{sec:normalization}
In single-cell data analysis, a crucial preprocessing step involves the {\it normalization} of the raw count matrix of the gene expression of each cell.
The variance of the expression level of a gene through cells should only depends on the gene.
The ``truly'' differentially expressed genes should exhibit high variance across cells, while others must have quite the same expression level.
Indeed, as discussed in \cite{Luecken2019}, differences among gene expressions might be due to sampling effects, the sequencing machine and its setting, and the variability among cell types.
Normalization aims to bring out the true differences among gene expressions.
In the literature, there are several normalization methods, and it is unclear which approach is the best.
%
In our work, we apply the {\it standard library size normalization in log-space}.
%
%
This normalization is competitive with other approaches (for an empirical survey, see \cite{lytal2020normalization}), and it is considered as best-practice in several recent SC-RNA Sequencing data analysis tutorials \cite{seow2020single}.
%

Given a gene expression profile $\mu$, the  {\it standard library size normalization in log-space} is defined as
\begin{equation}\label{std:norm}
    \hat \mu_i = \log \left(1 + \tau \times \frac{\mu_i}{\sum_{j \in \mathcal{G}} \mu_{j}}\right),
    \quad \forall i \in \mathcal{G},
\end{equation}
where the parameter $\tau$ depends on the cell sequencing machine used to collect the raw count matrix.
For instance, the authors of \cite{Nitzan2019} set $\tau = 10^5$, while
in \cite{van_galen_single-cell_2019,seow2020single}
the authors set $\tau=10^4$. In our case, we set $\tau=10^4$.

Notice that the normalization \eqref{std:norm} is conceptually different from \eqref{sumone}.
In \eqref{std:norm}, the normalization mitigates the effects of measurement errors and/or distortions.
In \eqref{sumone}, the normalization rescales the weights to make them sum up to 1, so as to permit a probability interpretation of the data.

\subsection{Combining Gene Profiles and Gene Embedding.}
In the ground distance \eqref{cost:g2v}, we implicitly assume to have an embedding for every gene in $\mathcal{G}$.
Unfortunately, this is not the case.
For instance, if we consider the AML dataset (i.e, GSE116256), we have the expression level of $27\,836$ genes, while the {\tt gene2vec} embedding we are using has the embedding only for $24\,447$ genes, which, in addition, it is not a subset of the genes expressed in GSE116256.
In practice, we are missing the embedding for $7\,523$ genes out of $27\,836$.

To reduce the impact of those missing gene embedding, we have looked for genes synonyms with the following method: first, we use \textit{genenames} to download a database of synonyms that, for each gene, provides all the possible names available in the literature \cite{GeneNames}.
Then, for each gene in \cite{van2019single}, we replace the names we got in the \textit{gene2vec} embedding with its synonyms in the \textit{genenames} dataset.
In this way, we are able to figure out that, for some genes, we have the actual embedded vector but under a different name.
For instance, in the case of AML, we retrieve an embedding for 108 more genes, and, hence, we have in total the embedding of $20\,484$ genes.
As future work, we could try to generate an embedding for every gene in $\mathcal{G}$.

\section{Computational Results}
\label{sec:comp}

The main goal of our computational tests is to compare the quality of different similarity measures between pairs of single-cells by using a $k$-Nearest Neighbour classifier ($k$-NN).
We use $k$-NN because its performances are strictly related to the quality of the metric used to look for {\it neighbours} \cite{Pele2009}.

For the computational tests, we have designed two types of classification tasks: the first task is designed for classifying the cells belonging to the same patient ({\it intrapatient}). The second task is to classify cells belonging to different patients ({\it interpatients}).
The motivation of the intrapatient task is to consider the case where a huge number of gene expression profiles of a single patient is available, but due to limited time and high cost of the (semi-manual) classification procedure, we can classify in vitro only a small fraction of cells.
The interpatients task has the objective of classifying the cells of a single patient, using as reference a given database of cells collected a priori and correctly classified.
Unfortunately, the interpatients task is very challenging because the gene expression profiles collected with different devices and different settings can yield very different results.
In both cases, if the automated classification were very accurate, the costly semi-manual classification procedure could be completely replaced by an automated procedure, reducing the cost and the time required by this type of medical analysis.

The metrics we use inside the $k$-NN algorithm are:
\begin{itemize}
  \item $d_E$: The Euclidean distance  defined in \eqref{dist:eucl}.
  \item $d_P$: The Pearson correlation score defined in \eqref{dist:pears}.
  \item $\mbox{GMD}_{\mbox{g2v}}$: The distance given by \eqref{bgmd:eq1}--\eqref{bgmd:ver} with the {\tt gene2vec} ground distance \eqref{cost:g2v}.
  \item $\mbox{GMD}_P$: The distance given by \eqref{bgmd:eq1}--\eqref{bgmd:ver} with the Pearson ground distance \eqref{cost:pearson}.
\end{itemize}
\noindent Table \ref{tab:runtime} reports the mean runtime for computing the distance between a pair of cells using the different metrics. For GMD, the mean runtime is the same for the two ground distances \eqref{cost:g2v} and \eqref{cost:pearson},
and is clearly larger than the runtime of the other two similarity measures.

In the following paragraphs, we detail the implementation of our algorithms, and we present our computational results.


\begin{table}[t]
\caption{Average of runtime to compute the different similarity measures among cells. The total runtime is given in hh:mm:ss time format, the mean pairwise runtime is given in milliseconds.\label{tab:runtime}}
\vskip 0.2in
\begin{center}
\begin{small}
\begin{sc}
\begin{tabular}{lcc}
\hline
& \multicolumn{2}{c}{Mean Runtime} \\
Metric  & Total  & Pairwise Distance  \\
\hline
Euclidean & 00:04:11 & 0.1 ms\\
Pearson & 00:04:09 & 0.1 ms \\
GMD &  04:46:39 & 8.6 ms \\
\hline
\end{tabular}
\end{sc}
\end{small}
\end{center}
\vskip -0.1in
\end{table}

\subsection{Implementation Details}
We have implemented in C++ a Network Simplex algorithm for the exact computation of the Gene Mover's Distance.
The algorithm is a fork the COIN-OR Lemon Graph \cite{Kovacs2015}, customized to solve uncapacitated transportation problems.
While the computation of the distance between a given pair of cells is sequential, we have parallelized on a multi-core processor the computation of the entries of the GMD matrix.

Table \ref{tab:runtime} shows the average runtime for computing the exact distances for the considered metrics, on a dataset with $1000$ cells, on a single node of a cluster having an INTEL CPU with 32 physical cores.
We report the average running time to compute a single pairwise distance, and the overall running time for computing all the exact distances to run the $k$-NN algorithm.
As expected from the literature, the computation of the GMD distance is very time-consuming.
However, in this medical context of critical human diseases, the major concern is the quality of the results, more than the running time for obtaining those results.

The classification algorithm is implemented in {\tt Python} using the {\tt scikit-learn} library \cite{scikitlearn}.
First, we precompute the full distance matrices in C++, for all the four similarity metrics.
Second, we parse those distance matrices in Python, and then we use {\tt KNeighborsClassifier} algorithm implemented in {\tt scikit-learn}.
The results on the intrapatient test are cross-validated using repeated stratified cross-validation, using 5 folds and 20 repetitions for each fold, via the {\tt RepeatedStratifiedKFold} cross validator.
During the experiments of $K$-fold validation, we have collected all the results for the following prediction scores: Accuracy and the Matthews Correlation Coefficient (MCC).
We tune the number of neighbours $k$ in each dataset by selecting the patient with fewer cells, and we looked at the performances for different values of $k$.
After cross-validation, we set $k=5$ for the AML dataset, and $k=3$ for the pancreas and brain dataset.

All our codes will be available on GitHub, in case of acceptance of this paper.

\subsection{Results for AML dataset}
For the AML dataset, we perform a binary classification to classify each single-cell as either {\it normal} or {\it malignant}.
Table \ref{tab:patientsAML} describes the 20 single-cell sequencing experiments that we selected to design our classification instances.
The first column in the table denotes the name of the single-cell RNA sequencing experiments, where the name of the experiment includes the identifier of a single patient: note that the same patient appears twice, first at day zero (AML329-D0) and later at day 20 (AML329-D20).
In our test, we treat each experiment independently, regardless of the patient they refer to.
The second, third, and fourth columns of Table \ref{tab:patientsAML} reports the total number of cells for each experiment, the total number of malignant cells, and the percentage of malignant cells.

\begin{table}[t!]
\caption{Experiments from GSE116256 selected to design our classification instances: name of the experiment, number of cells, number of malignant cells, and percentage of malignant cells \cite{van_galen_single-cell_2019}. We selected only cells with a label specified.}
\label{tab:patientsAML}
\vskip 0.15in
\begin{center}
\begin{small}
\begin{sc}
\begin{tabular}{lrrr}
\hline
Experiment & Cells & Malignant & $\%$ Malig. \\
\hline
AML419A-D0 & 1189 & 1068 & 89.8\% \\
AML916-D0 & 933 & 775 & 83.1\% \\
AML1012-D0 & 1136 & 856 & 75.3\% \\
AML475-D0 & 423 & 308 & 72.8\% \\
AML328-D0 & 1094 & 693 & 63.3\% \\
AML210A-D0 & 748 & 464 & 62.0\% \\
AML329-D0 & 525 & 253 & 48.2\% \\
AML329-D20 & 953 & 259 & 27.2\% \\
AML420B-D0 & 485 & 100 & 20.6\% \\
AML328-D171 & 1402 & 161 & 11.5\% \\
\hline
BM3 & 643 & 0 & 0.00\% \\
BM4 & 3738 & 0 & 0.00\% \\
AML371-D0 & 56 & 0 & 0.00\% \\
AML556-D15 & 1203 & 0 & 0.00\% \\
AML420B-D14 & 1282 & 1 & 0.08\% \\
AML328-D29 & 1880 & 1145 & 60.90\% \\
AML707B-D0 & 1586 & 1370 & 86.38\% \\
AML556-D0 & 2328 & 2062 & 88.57\% \\
AML921A-D0 & 3813 & 3378 & 88.59\% \\
AML870-D0 & 345 & 314 & 91.01\% \\
\hline
\end{tabular}
\end{sc}
\end{small}
\end{center}
\vskip -0.1in
\end{table}

\paragraph{Intrapatient classification.} For the intrapatient tests, we have selected 10 experiments provided in the original dataset.
These 10 experiments have a number of cells that range between 400 and $1\,400$, with a percentage of malignant cells ranging from $11.48\%$ to $89.82\%$.
Note that in the dataset each single-cell is labeled as {\tt normal=0} or {\tt malignant=1}.

For the binary classification, we proceed as follows:
(a) we run a $k$-NN with training and test sets randomly chosen by using the {\tt sklearn.model\_selection.train\_test\_split} function with a fixed seed;
(b) we compute the prediction scores MCC and Accuracy (ACC);
(c) we repeat 10 times steps (a)--(b) by changing the seed, that is, by using a different split of the training and test set.
Since the AML dataset is unbalanced between the number of normal and malignant cells, we focus our discussion on the MCC values:
as discussed in \cite{chicco2020advantages,hand2018note}, MCC is likely the best prediction score when the dataset is unbalanced.

\paragraph{Interpatients classification.} For the interpatients tests, we consider 10 additional experiments which correspond to the second half of Table \ref{tab:patientsAML}, from experiment BM3 to AML870-D0.
We selected those 10 experiments because they correspond to five experiments where cells are mostly normal, and to five experiments where cells are mostly malignant.
We remark that there is no overlap between the experiments used for the reference and the test sets.

Then, we perform the binary classification as follows: (1) We build a dataset by randomly select 500 cells from the healthy patients from the second half of Table~\ref{tab:patientsAML} and 500 from the AML ones (from now on: TeP).
Then we append 100 randomly selected cells for each patient of the first half of Table \ref{tab:patientsAML} (from now on: TrP).
Thus, we get a dataset with 1500 cells. (2) We pick all the patients of the TeP and we run $k$-NN by using as a training set all the patients of TrP.
For the evaluation, we use the same scores of the intrapatient instance.
We create 5 different dataset by randomly selecting the cells, and we repeat step (2) for each of them.

\begin{figure}[t!]
\begin{center}
\centerline{\includegraphics[width=\columnwidth]{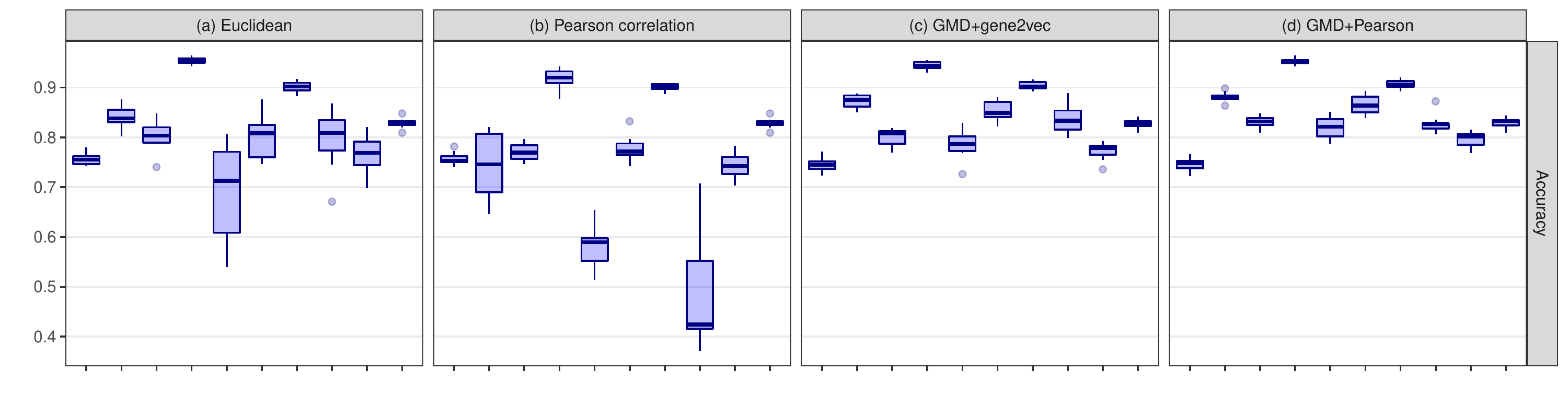}}
\centerline{\includegraphics[width=\columnwidth]{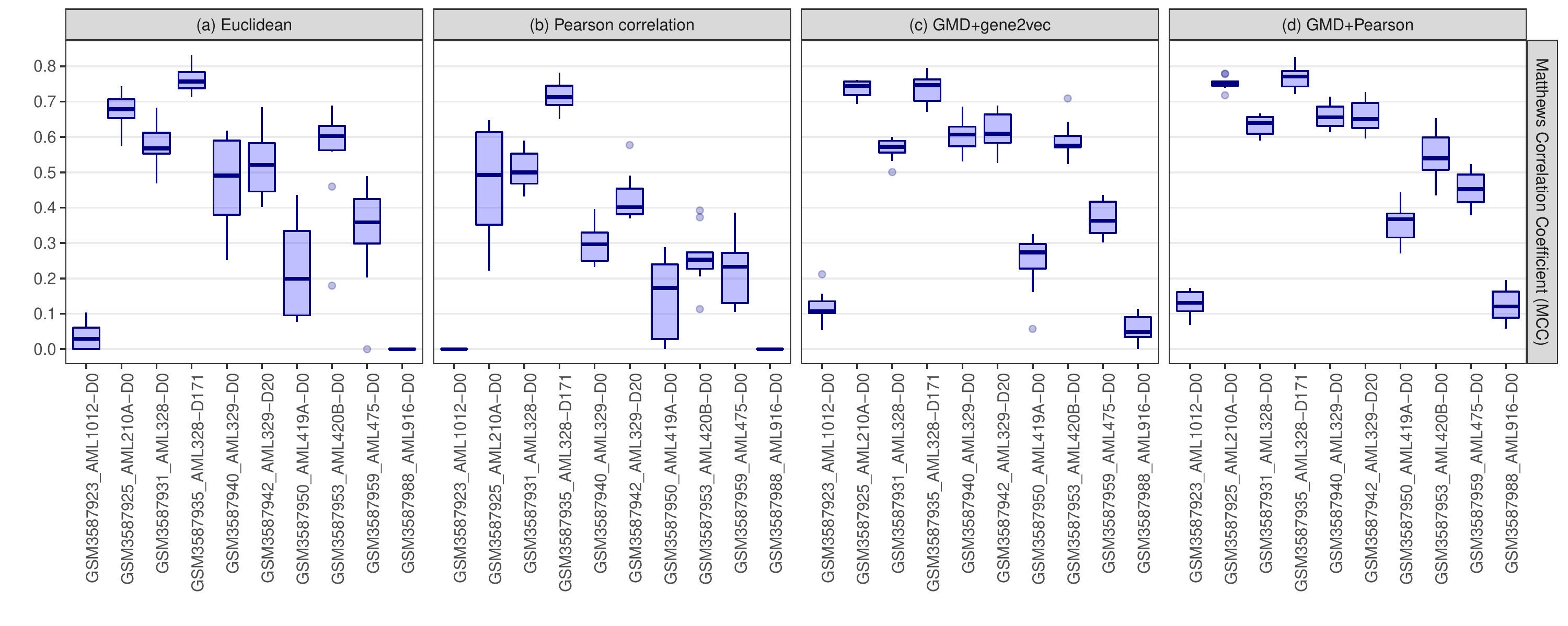}}
\caption{Accuracy and Matthews Correlation Coefficient (MCC) prediction scores for the {\bf intrapatient} classification instances for the AML dataset.}
\label{fig:intra}
\end{center}
\end{figure}

\begin{figure}[t!]
\begin{center}
\centerline{\includegraphics[width=\columnwidth]{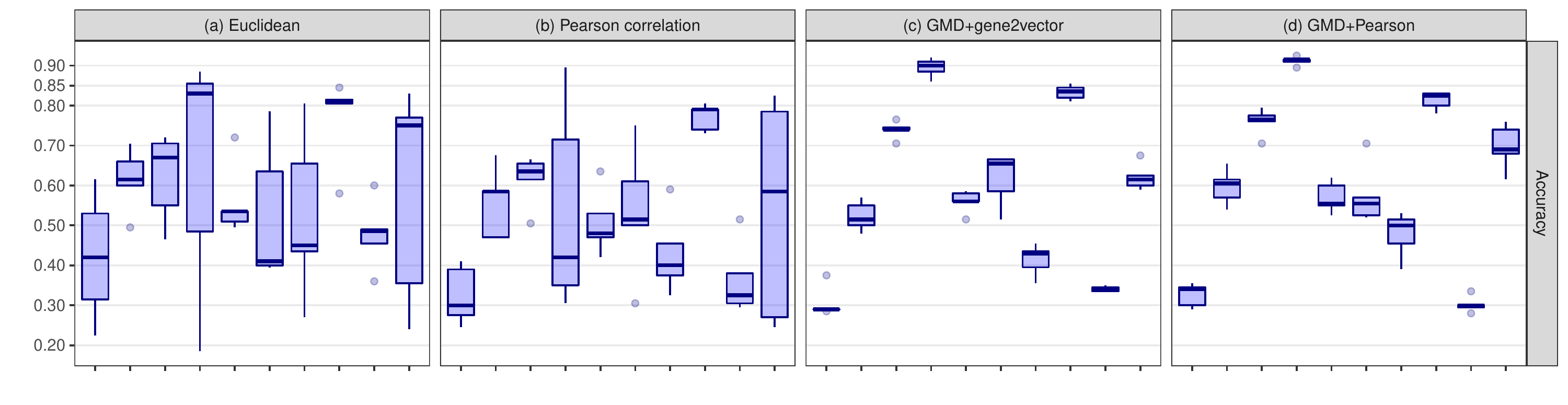}}
\centerline{\includegraphics[width=\columnwidth]{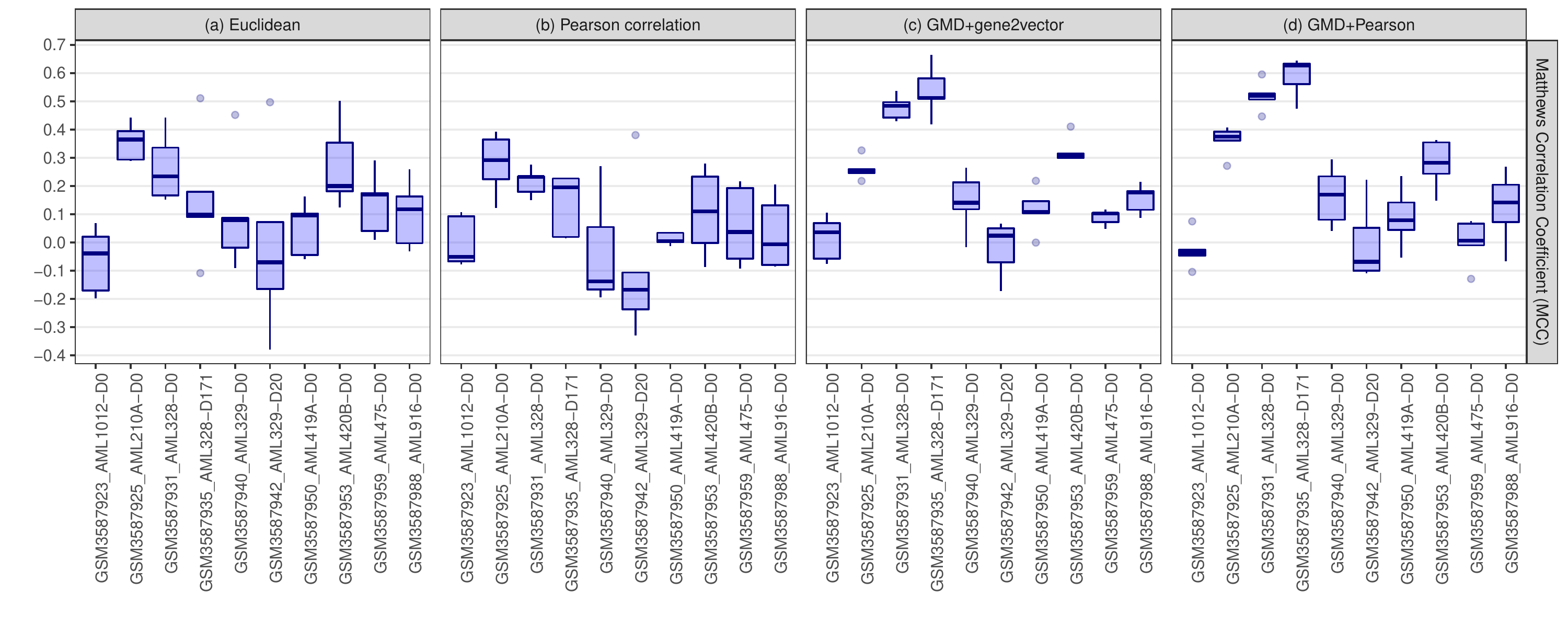}}
\caption{Accuracy and Matthews Correlation Coefficient (MCC) prediction scores for the {\bf interpatients} classification instances for the AML dataset GSE116256.}
\label{fig:inter}
\end{center}
\end{figure}

\begin{table}[t!]
\caption{Comparison of pair of classifiers for statistical significance for the MCC score using the $t$-test.
The 2nd and 3rd columns compare the results of $k$-NN with $\mbox{GMD}_{\mbox{g2v}}$ versus the Euclidean distance $d_E$ and the Pearson correlation $d_P$.
The 4th and 5th columns compare $\mbox{GMD}_{P}$ with the other two similarity functions.
The results are statistically significant only when $p$-value is smaller than $0.05$.}
\label{tab:ttest}
\begin{center}
\begin{small}
\begin{tabular}{lcccc}
\hline
 & \multicolumn{2}{c}{$\mbox{GMD}_{\mbox{g2v}}$ vs. $d_E$} & \multicolumn{2}{c}{$\mbox{GMD}_{\mbox{g2v}}$ vs.  $d_P$} \\
Experiment & $p$-value &  & $p$-value &  \\
\hline
AML1012-D0 &  0.000& $\surd$   & 0.000  & $\surd$\\
AML210A-D0 &  0.001& $\surd$   & 0.000  & $\surd$\\
AML328-D0 &   0.514&  $\times$ & 0.010  & $\surd$\\
AML328-D171 &   0.149& $\times$   & 0.259  & $\times$\\
AML329-D0 &   0.007& $\surd$   & 0.000  & $\surd$\\
AML329-D20 &  0.014& $\surd$   & 0.000  & $\surd$\\
AML419A-D0 &  0.602& $\times$   & 0.035  & $\surd$\\
AML420B-D0 &  0.471& $\times$   & 0.000  & $\surd$\\
AML475-D0 &   0.466&  $\times$ & 0.000  & $\surd$\\
AML916-D0 &   0.000& $\surd$   & 0.001  & $\surd$\\
\hline
\\
 & \multicolumn{2}{c}{$\mbox{GMD}_{P}$ vs.  $d_E$} & \multicolumn{2}{c}{$\mbox{GMD}_{P}$ vs.  $d_P$} \\
Experiment & $p$-value &  & $p$-value &  \\
\hline
AML1012-D0 &  0.000& $\surd$   & 0.000  & $\surd$\\
AML210A-D0 &  0.000& $\surd$   & 0.000  & $\surd$\\
AML328-D0 &   0.025& $\surd$   & 0.000  & $\surd$\\
AML328-D171 & 0.830& $\times$  & 0.003  & $\surd$\\
AML329-D0 &   0.000& $\surd$   & 0.000  & $\surd$\\
AML329-D20 &  0.000& $\surd$   & 0.000  & $\surd$\\
AML419A-D0 &  0.008& $\surd$   & 0.000  & $\surd$\\
AML420B-D0 &  0.900& $\times$  & 0.000  & $\surd$\\
AML475-D0 &   0.026& $\surd$   & 0.000  & $\surd$\\
AML916-D0 &   0.000& $\surd$   & 0.000  & $\surd$\\
\hline
\end{tabular}
\end{small}
\end{center}
\vskip -0.1in
\end{table}

\paragraph{Results.} The computational results on the AML dataset are reported in Figure~\ref{fig:intra} for the intrapatient task and in Figure~\ref{fig:inter} for the interpatients task.
While any single similarity measure dominates the others in terms of prediction scores, the striking result is that the GMDs are always very precise, having a small interquartile range.
We can observe that on the intrapatient task of Figure~\ref{fig:intra} the two GMDs obtain good Accuracy (always larger than 0.7) and discrete MCC prediction scores.

On the interpatients tasks of Figure~\ref{fig:inter} the prediction scores are more heterogeneous, and indeed the task of correctly classify every single-cell as normal or malignant by only using gene expression profiles coming from other patients and/or experiments is very challenging.
However, we observe a great variability of the two prediction scores over every single experiment.
For instance, the patient AML328 looks quite easy to classify both at day D0 and at day D171, since we have a good accuracy score and high MCC value.
On the contrary, the patient AML1012 at day D0 yields very poor results, regardless of the metric used in the $k$-NN classifier.
However, the cells of the patient AML1012 were classified with good results in the interpatients instances.

The performance of the classifiers based on the different metrics are assessed in terms of statistical significance by performing a two-sided $t$-test for the MCC score for the null hypothesis that 2 independent samples have identical expected values.
We use the function ${\tt ttest\_ind }$ of the ${\tt python \ SciPy}$ package \cite{2020SciPy-NMeth}.
This test assumes that the populations have identical variances by default.

We report our results in Table \ref{tab:ttest}.
In around half of the instances, the $t$-test rejects the null hypothesis and confirms that the GMD distances yields, on average, better results for the MCC prediction score regardless of the cost function.
Furthermore, even when the null hypothesis is accepted, for instance in Patient AML328 at day 171, the $k$-NN classifiers with the GMD metrics outperforms the same classifiers with both Euclidean distance and Pearson correlation score.
%

\subsection{Results for the Pancreas dataset}
For this dataset, we perform a multi-class classification using the class type of each cell.
Due to the reduced size of this dataset, we use all four patients for the intrapatient task, and we run the classification using the same approach of the AML dataset.
In the interpatients task, we generate six datasets by considering every possible combination of patients (two for training, two for testing).
In addition, we limit the dataset to contain no more than 80 cells for each class, in both the training and test set.
As a consequence, the average dimension of each dataset is of $1400$ cells.
Then, we perform step (2) as in the previous subsection, and we collect the prediction scores for Accuracy and MCC.

\begin{figure}[t!]
\begin{center}
\centerline{\includegraphics[width=\columnwidth]{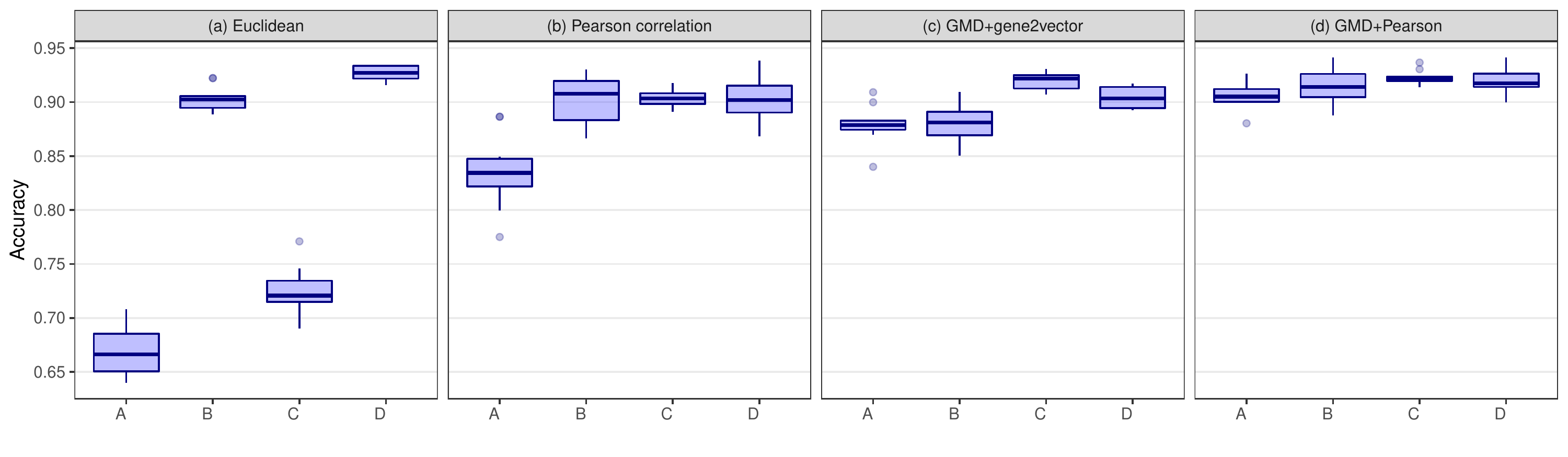}}
\centerline{\includegraphics[width=\columnwidth]{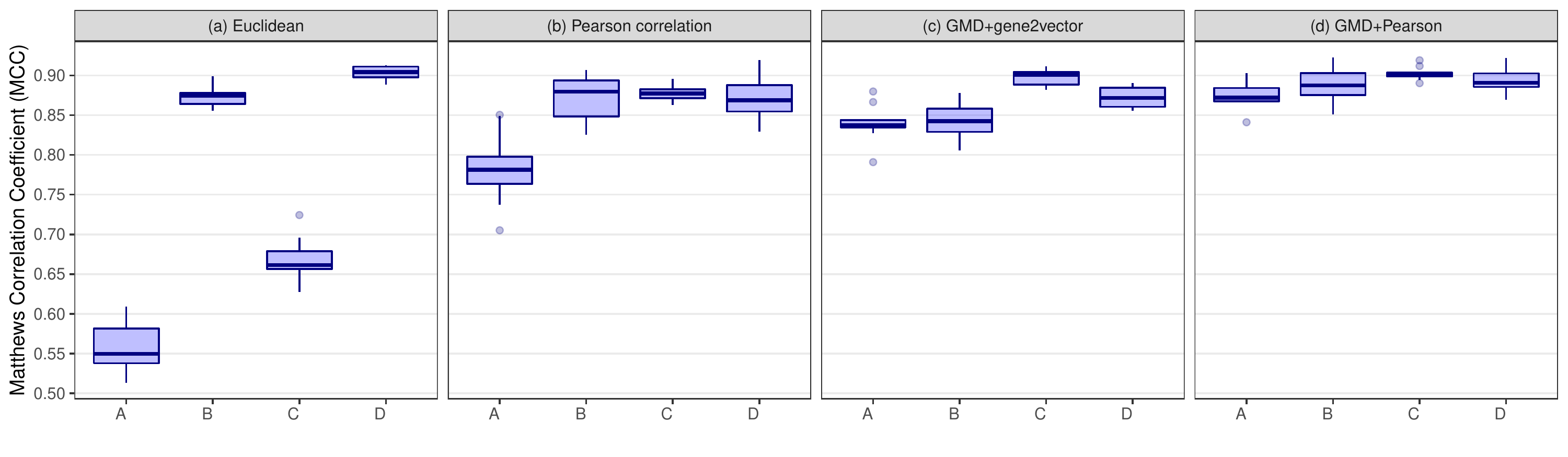}}
\caption{Accuracy and Matthews Correlation Coefficient (MCC) prediction scores for the {\bf intrapatient} classification instances for the Pancreas dataset GSE84133.}
\label{fig:pancreas}
\end{center}
\end{figure}

Figure~\ref{fig:pancreas} shows the computational results for the intrapatient tests, where both the GMDs achieve better results than the Pearson and Euclidean metrics.
Differently than for the binary classification of the AML dataset, the Pearson correlation score performs very well, and the Euclidean very badly.
%
%
In addition, as noted for the AML dataset, the boxplots generated by the GMD metrics exhibit a very small interquartile  range of the prediction scores, meaning that the GMDs are very precise.

Table~\ref{tab:pancreas} reports the results of the $t$-tests, which are similar to the results obtained on the AML dataset.

\begin{table}[t!]
\caption{Comparison of pair of classifiers for statistical significance for the MCC score using the $t$-test for the GSE84133 dataset (pancreatic cells dataset), \emph{intra-patient} instance.
The 2nd and 3rd columns compare the results of $k$-NN with $\mbox{GMD}_{\mbox{g2v}}$ versus the Euclidean distance $d_E$ and the Pearson correlation $d_P$.
The 4th and 5th columns compare $\mbox{GMD}_{P}$ with the other two similarity functions.
The results are statistically significant only when $p$-value is smaller than $0.05$. \label{tab:pancreas}}
\begin{center}
\begin{small}
\begin{tabular}{lcccc}
 \hline
 & \multicolumn{2}{c}{$\mbox{GMD}_{\mbox{g2v}}$ vs. $d_E$} & \multicolumn{2}{c}{$\mbox{GMD}_{\mbox{g2v}}$ vs.  $d_P$} \\
Experiment & $p$-value &  & $p$-value &  \\
\hline
A & 0.000         & $\surd$ & 0.002 & $\surd$       \\
B & 0.001         & $\surd$ & 0.025 & $\surd$       \\
C & 0.000         & $\surd$ & 0.000 & $\surd$       \\
D & 0.000         & $\surd$ & 0.896 & $\times$       \\
 \hline
\\
 & \multicolumn{2}{c}{$\mbox{GMD}_{P}$ vs.  $d_E$} & \multicolumn{2}{c}{$\mbox{GMD}_{P}$ vs.  $d_P$} \\
Experiment & $p$-value &  & $p$-value &  \\
\hline
A  & 0.000         & $\surd$ & 0.000  & $\surd$    \\
B  & 0.124         & $\times$ & 0.180 & $\times$      \\
C  & 0.000         & $\surd$ & 0.000  & $\surd$    \\
D  & 0.148        & $\times$ & 0.031 & $\surd$ \\
\hline
\end{tabular}
\end{small}
\end{center}
\vskip -0.1in
\end{table}

\subsection{Results for the Brain dataset}
The brain dataset is interesting because it has a small ratio between cells and class types and it contains several patients (see Table~\ref{tab:patientsAML}).
We consider it meaningless to perform an intrapatient instance as we have noticed that the maximum amount of cells per patient is $70$.
This is a good example of when the interpatients instance may result meaningful, as small samples need a pre-determined and independent training set to be classified properly.
Thus, we build only three different datasets: (a) fetal brain's cells, (b) adult brain cells, (c) fetal+adult brain cells.
Then we run $k$-NN with training and test sets randomly chosen by using the {\verb|sklearn.model_selection.train_test_split|} function with a fixed seed.
Finally, we change the seed 10 times, and each time we collect the same prediction scores as before.

\begin{figure}[t!]
\begin{center}
\centerline{\includegraphics[width=\columnwidth]{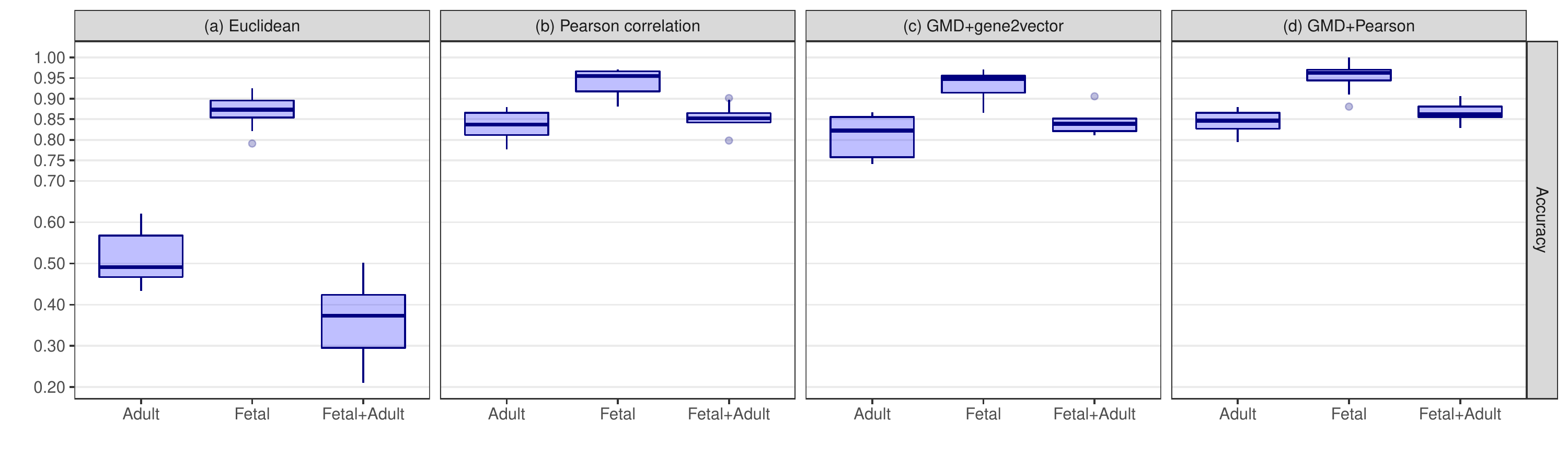}}
\centerline{\includegraphics[width=\columnwidth]{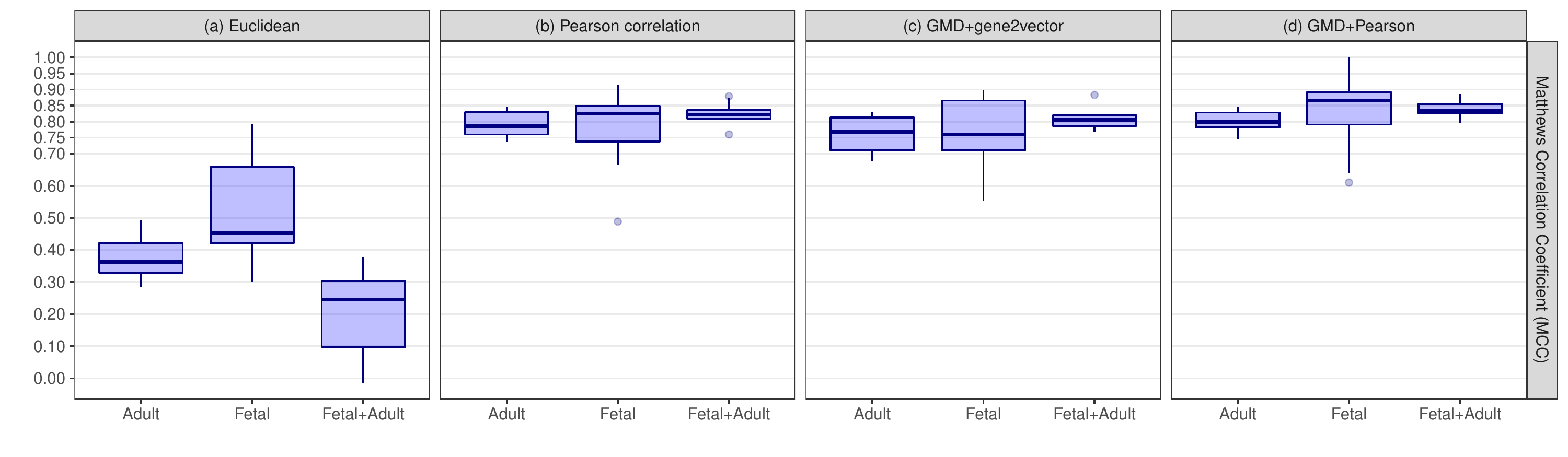}}
\caption{Accuracy and Matthews Correlation Coefficient (MCC) prediction scores for the {\bf interpatients} classification instances for the brain dataset GSE67835.}
\label{fig:brain}
\end{center}
\end{figure}

Figure~\ref{fig:brain} shows the results for the interpatients tests. In this case, the two similarity measures based on the Pearson coefficients achieve the best prediction scores.
Both measures have an accuracy larger than 0.8.
The Euclidean distance is clearly worse, while the GMD based on the {\tt gene2vector} embedding falls slightly behind the other two measures.

Table~\ref{tab:brain} reports the results of the $t$-tests for the brain dataset.
In this case, the $t$-test confirms that the GMDs are statistically more significant than the results of the Euclidean distance.
Moreover, the $t$-test highlight that, on the brain dataset, there is no statistically significant difference between the Pearson correlation score and the two GMDs.

\begin{table}[t!]
\caption{Comparison of pair of classifiers for statistical significance for the MCC score using the $t$-test for the GSE67835 dataset (brain cells dataset), \emph{inter-patients} instance.
The 2nd and 3rd columns compare the results of $k$-NN with $\mbox{GMD}_{\mbox{g2v}}$ versus the Euclidean distance $d_E$ and the Pearson correlation $d_P$.
The 4th and 5th columns compare $\mbox{GMD}_{P}$ with the other two similarity functions.
The results are statistically significant only when $p$-value is smaller than $0.05$.\label{tab:brain}}
\begin{center}
\begin{small}
\begin{tabular}{lcccc}
 \hline
 & \multicolumn{2}{c}{$\mbox{GMD}_{\mbox{g2v}}$ vs. $d_E$} & \multicolumn{2}{c}{$\mbox{GMD}_{\mbox{g2v}}$ vs.  $d_P$} \\
Experiment & $p$-value &  & $p$-value &  \\
\hline
Adult         &0.000  & $\surd$ & 0.154 & $\times$\\
Fetal         & 0.002  & $\surd$ & 0.667 &  $\times$\\
Adult + Fetal & 0.000  & $\surd$ & 0.241 &  $\times$\\
  \hline
\\
 & \multicolumn{2}{c}{$\mbox{GMD}_{P}$ vs.  $d_E$} & \multicolumn{2}{c}{$\mbox{GMD}_{P}$ vs.  $d_P$} \\
Experiment & $p$-value &  & $p$-value &  \\
\hline
Adult         & 0.000  & $\surd$ & 0.606 & $\times$  \\
Fetal         & 0.000  & $\surd$ & 0.358 & $\times$  \\
Adult + Fetal & 0.000  & $\surd$ & 0.266 & $\times$   \\
 \hline
\end{tabular}
\end{small}
\end{center}
\vskip -0.1in
\end{table}

%




%

\section{Conclusions}
\label{sec:concl}

In this paper, we have proposed an application of Computational Optimal Transport to measure the similarity of single-cell expression profiles: the Gene Mover's Distance.
We have run extensive computational tests to compare the proposed distance with other similarity metrics using three different biological datasets.
For the classification task, we have used a $k$-NN algorithm because its main component is the similarity function used to compare the items.

Our computational results confirm that to classify a single-cell as either normal or malignant by using only the gene expression profile is very challenging.
However, we have shown that the two most used similarity score in single-cell data analysis \cite{Luecken2019}, namely the Euclidean and the Pearson similarity measures exhibit a very high variance in the prediction scores, and, hence, they should be used with care.
On the contrary, the GMD distances give very good results with very small interquantile distance.
Between the {\tt gene2vec} and the Pearson cost functions, the latter seems to perform better.
This might be due to the missing embedded vectors for a number of genes: Indeed, while the GMD induced by the Pearson ground distance, by construction, considers all the genes present in the datasets, the GMD induced by {\tt gene2vec} only considers genes for which \cite{Du2019} was able to find an embedding.

The results of this work raise several important research questions for future works.
Firstly, regarding the GMD distance based on the {\tt gene2vec} embedding, we are missing the embedded vector of a few thousand of genes.
Hence, it would be desirable to compute a gene embedding specifically designed for all the tissues (blood, pancreas, and brain) we have considered.
These new embedding could be derived with the same approach used in \cite{Du2019}.
Secondly, in terms of Optimal Transport distances, there are other formulations that could be considered, as for instance the entropic unbalanced optimal approach (e.g., see \cite{liero2018optimal}).
%

%
%
%

\section*{Acknowledgements}
This research was partially supported by the Italian Ministry of Education, University and Research (MIUR):
Dipartimenti di Eccellenza Program (2018--2022) - Dept. of Mathematics ``F. Casorati'', University of Pavia.
The PhD scholarship of Andrea Codegoni is sponsored by Sea Vision Srl.


\end{document}